\pdfoutput=1
\documentclass{article}

\usepackage{arxiv}

\usepackage[utf8]{inputenc} % allow utf-8 input
\usepackage[T1]{fontenc}    % use 8-bit T1 fonts
\usepackage{hyperref}       % hyperlinks
\usepackage{url}            % simple URL typesetting
\usepackage{booktabs}       % professional-quality tables
\usepackage{amsfonts}       % blackboard math symbols
\usepackage{nicefrac}       % compact symbols for 1/2, etc.
\usepackage{microtype}      % microtypography
\usepackage{lipsum}

\usepackage{amsmath} %% JK added this May 6

\title{Dyadic Reciprocity as a Function of Covariates}

\author{
  Jeremy Koster\\
  Department of Human Behavior, Ecology, and Culture\\
  Max Planck Institute for Evolutionary Anthropology\\
  Leipzig, Germany \\
  \texttt{jeremy\_koster@eva.mpg.de} 
}

\begin{document}
\maketitle

\begin{abstract}
Reciprocity in dyadic interactions is common and a topic of interest across disciplines. In some cases, reciprocity may be expected to be more or less prevalent among certain kinds of dyads. In response to interest among researchers in estimating dyadic reciprocity as a function of covariates, this paper proposes an extension to the multilevel Social Relations Model. The outcome variable is assumed to be a binomial proportion, as is commonly encountered in observational and archival research. The approach draws on principles of multilevel modeling to implement random intercepts and slopes that vary among dyads. The corresponding variance function permits the computation of a dyadic reciprocity correlation. The modeling approach can potentially be integrated with other statistical models in the field of social network analysis.
\end{abstract}

% keywords can be removed
\keywords{Social Network Analysis \and Dyadic Reciprocity \and Social Relations Model}

\section{Introduction}

In recent decades, methods and concepts from social network analysis have been incorporated into diverse academic disciplines, ranging from sociology to ecology to political science and physics \cite{borgatti2009network, sih2009social}. Although social network analysis is often associated with survey research, observational and archival data present opportunities to study dyadic behavior, arguably without the biases that characterize survey research with informants \cite{bernard1984problem,kashy1990you, lewis2008tastes, quintane2011matter, aven2015paradox, aven2019valley}. In many settings, particularly studies of animal behavior, observational research may represent the only viable option for obtaining data on social relationships. Often, the datasets from these studies are characterized by observations of events over a period of time. For instance, ecologists may document whether individual \textit{i} helped individual \textit{j} during an observation.

For ecologists, there is often substantive interest in the extent of dyadic reciprocity between individuals in a population. This interest relates in part to theoretical models of cooperation, where contingent reciprocity represents one possible path toward helping behavior among individuals \cite{cheney2010contingent}. By extension, researchers have considered the extent to which dyadic reciprocity may vary as a function of another variable. For instance, individuals who share a kinship tie may exhibit greater reciprocity than unrelated individuals \cite{allen2008reciprocal}.

The statistical analysis of relational data poses well-known challenges, however, and multiple statistical approaches have been advanced to address the structural dependencies that typify these data \cite{snijders2011statistical}. One approach is the multilevel Social Relations Model (SRM), which employs random effects to partition data into separate giving, receiving, and relational components 
\cite{snijders1999social,koster2019statistical}. Originally developed for continuous responses, the SRM can be adapted for discrete responses, such as binary and count outcomes \cite{koster2018effects,koster2014food,loncke2019social}.

Conventional applications of the SRM, however, assume that dyadic reciprocity is constant across all dyads in the sample. In other words, the conventional SRM cannot be used to assess the extent to which dyadic reciprocity varies as a function of a covariate. Here I advance an extension of the multilevel SRM that incorporates varying slopes to permit such analyses. The model assumes that dyads have been observed multiple times and that the outcome variable is a binomial proportion.

\section{The Statistical Model}

Imagine an outcome variable, $y_{ij}$, in which the response reflects a binomial proportion where the numerator is the number of times that node \textit{i} directed a behavior toward node \textit{j} and the denominator represents the total number of opportunities for the behavior to have been directed. Further imagine a predictor variable, $x_{ij}$, that represents a dyadic characteristic, such as the degree of relatedness between individuals $i$ and $j$ \cite{koster2018family}.

A binomial regression model can then be specified:

\begin{equation}\label{eq1}
  \begin{gathered}
y_{ij} \sim \textrm{binomial} (n_{ij}, p_{ij}) \\
\textrm{logit}(p_{ij}) = \alpha + a_i + b_j + \beta (x_{ij}) + u_{|ij|} + v_{|ij|}(x_{ij}) + d_{ij} \\
  \end{gathered}
\end{equation}

where $\alpha$ represents the intercept, $a_i$ is a node-level random intercept for the sending node $i$, $b_j$ is an analogous random intercept for incoming events to $j$, $\beta (x_{ij})$ is a conventional fixed effect parameter for the predictor, $u_{|ij|}$ and $v_{|ij|}(x_{ij})$ are the respective random intercept and slope for dyad $ij$, and $d_{ij}$ is a parameter that captures possible overdispersion beyond the conditional expectation for the binomial distribution.

The random effects for sending and recipient nodes are assumed bivariate normally distributed with zero means and homogeneous covariance matrix
\begin{align*}
\left(
\begin{matrix} a_{i} \\ b_{i} 
\end{matrix}
\right)
&\sim \textrm{Normal} \left\lbrace \left( \begin{matrix}  0 \\ 0  \end{matrix} \right) , \left( \begin{matrix}  \sigma^{2}_a &  \\ \sigma_{ab} & \sigma^{2}_b  \end{matrix} \right)
\right\rbrace
\end{align*}
As in other applications of the SRM \cite{jorgensen2018using}, the correlation between the respective effects is known as the "generalized reciprocity correlation."

The random effects for the dyad-level intercepts and slopes are likewise assumed bivariate normally distributed with zero means and homogeneous covariance matrix
\begin{align*}
\left(
\begin{matrix} u_{|ij|} \\ v_{|ij|} 
\end{matrix}
\right)
&\sim \textrm{Normal} \left\lbrace \left( \begin{matrix}  0 \\ 0  \end{matrix} \right) , \left( \begin{matrix}  \sigma^{2}_u &  \\ \sigma_{uv} & \sigma^{2}_v  \end{matrix} \right)
\right\rbrace
\end{align*}
Importantly, the notation of $|ij|$ is an indicator for a symmetric dyadic relationship. That is, within each dyad, the relationship from $i$ to $j$ and the relationship from $j$ to $i$ share the same index \cite{koster2015multilevel}.

The inclusion of the parameter for additive overdispersion follows Browne et al. \cite{browne2005variance}. As in standard applications of multilevel modeling, these effects are assumed to be normally distributed:
\begin{equation*}\
d_{ij} \sim \textrm{Normal}(0, \sigma^{2}_d)
\end{equation*}
There may be binomial data that exhibit minimal overdispersion, in which case this parameter could potentially be omitted.

\section{Estimating Dyadic Reciprocity}
As noted, this approach estimates symmetric dyad effects. When this approach is used in a conventional SRM for continuous data, Snijders and Kenny \cite{snijders1999social} show that dyadic reciprocity, $\rho$, can be estimated as:
\begin{equation*}
\rho = \sigma^{2}_u / (\sigma^{2}_u + \sigma^{2}_e)
\end{equation*}
where $\sigma^{2}_u$ is the variance of the symmetric dyad effects and $\sigma^{2}_e$ is the residual variance (implicitly corresponding to the directed dyadic observations, which are the unit of analysis). This parameterization constrains dyadic reciprocity to be positively correlated, which may be a reasonable assumption in many cases.

The same logic can be used to estimate dyadic reciprocity in a binomial SRM. However, these models lack the constant residual variance of Gaussian regression models. As an alternative, latent parameterizations of binomial models may assume that the corresponding variance is $\pi^2/3$, or 3.29 \cite{snijders2011multilevel}. This quantity can be substituted into the denominator along with the variance for overdispersion effects.

Unlike a conventional SRM with symmetric dyadic effects, the dyadic variance in the above model is not constant. Rather, it varies as a function of the predictor variable, $x_{ij}$. Snijders \cite{snijders2011multilevel} notes that in multilevel models with random slopes, the variance follows a quadratic function of the predictor:
\begin{equation*}
\sigma^{2}_u + 2\sigma_{uv}x_{ij} + \sigma^{2}_vx^2_{ij}
\end{equation*}

This expression can take the place of the dyadic variance, yielding a corresponding calculation for dyadic reciprocity:
\begin{equation*}
\rho = {\frac 
{\sigma^{2}_u + 2\sigma_{uv}x_{ij} + \sigma^{2}_vx^2_{ij}}
{\sigma^{2}_u + 2\sigma_{uv}x_{ij} + \sigma^{2}_vx^2_{ij} + \sigma^{2}_d + 3.29}
}
\end{equation*}

Once the parameters from the statistical model have been estimated, the dyadic reciprocity correlation can be calculated for different values of $x$.

\section{Discussion}
This paper introduces a binomial Social Relations Model that permits the estimation of dyadic reciprocity as a function of a predictor variable. This is accomplished via the parameterization of random effects that correspond to a varying intercept and a varying slope in conventional applications of multilevel modeling \cite{goldstein2011multilevel}. The dyadic variance can then vary as a function of the predictor variable and compared against the residual variance to calculate reciprocity. The present model could be expanded further with the inclusion of other covariates and parameters.

The model here assumes that the outcome is composed of binomial proportions. Such data may be common among researchers who use observational or archival data. In principle, the data could be disaggregated so that the response represent a single binary observation, in which case a comparable random effects structure above could be adapted while permitting analogous estimates of dyadic reciprocity \cite{mcelreath2020statistical}.

The use of the SRM often assumes a "round robin" design in which the dyadic relationships among the individuals or nodes are fully observed. In observational studies, however, individuals may not be simultaneously present. In experimental studies, occasionally a subset of dyadic interactions are precluded by the research design \cite{finkel2008speed, argote2018effects}. As with broader applications of the Social Relations Model to "block" designs \cite{back2010social}, the above parameterization could likewise be employed by omitting such unobserved dyads from the analysis.

Whereas the SRM partitions variance according to givers, receivers, and relational components, a number of researchers have observed that social networks often exhibit additional structural dependencies, such as transitivity and block structures \cite{krivitsky2009representing, sweet2018estimating, minhas2019inferential}. In this literature, dyadic reciprocity are often a secondary consideration relative to other aspects of network structure (cf. \cite{zijlstra2017regression}).\footnote{Other modeling approaches emphasize the temporal dynamics in longitudinal dyadic data, with relatively greater attention to reciprocity \cite{butts20084, stadtfeld2017interactions}.} It may be possible for these statistical approaches to network data to be modified to examine dyadic reciprocity in greater detail alongside other aspects of the data-generating process.

An advantage of the modeling approach above is that it builds on principles that are common in the multilevel modeling literature, namely random intercepts and slopes and the corresponding variance function. One consequence of this approach is that the model is highly parameterized, as two parameters are estimated for each dyad. Another consideration is that the dyadic reciprocity correlation is constrained to be positive. As an alternative, Koster et al. \cite{koster2015multilevel} suggested estimating the reciprocity correlation via a tanh link function that allows the dyadic variance and covariance to be modeled as a function of predictors \cite{leckie2014modeling}. In some cases, that latter approach could be more advantageously implemented in a Social Relations Model than the present alternative, particularly in contexts where negative reciprocity correlations might be expected.

\newpage

\bibliographystyle{unsrt}  
%\bibliography{references}

\begin{thebibliography}{10}

\bibitem{borgatti2009network}
Stephen~P Borgatti, Ajay Mehra, Daniel~J Brass, and Giuseppe Labianca.
\newblock Network analysis in the social sciences.
\newblock {\em Science}, 323(5916):892--895, 2009.

\bibitem{sih2009social}
Andrew Sih, Sean~F Hanser, and Katherine~A McHugh.
\newblock Social network theory: new insights and issues for behavioral
  ecologists.
\newblock {\em Behavioral Ecology and Sociobiology}, 63(7):975--988, 2009.

\bibitem{bernard1984problem}
H~Russell Bernard, Peter Killworth, David Kronenfeld, and Lee Sailer.
\newblock The problem of informant accuracy: The validity of retrospective
  data.
\newblock {\em Annual Review of Anthropology}, 13(1):495--517, 1984.

\bibitem{kashy1990you}
Deborah~A Kashy and David~A Kenny.
\newblock Do you know whom you were with a week ago friday? {A} re-analysis of
  the {Bernard, Killworth, and Sailer studies}.
\newblock {\em Social Psychology Quarterly}, pages 55--61, 1990.

\bibitem{lewis2008tastes}
Kevin Lewis, Jason Kaufman, Marco Gonzalez, Andreas Wimmer, and Nicholas
  Christakis.
\newblock Tastes, ties, and time: A new social network dataset using
  facebook.com.
\newblock {\em Social networks}, 30(4):330--342, 2008.

\bibitem{quintane2011matter}
Eric Quintane and Adam~M Kleinbaum.
\newblock {Matter over mind? E-mail data and the measurement of social
  networks}.
\newblock {\em Connections}, 13(1):22--46, 2011.

\bibitem{aven2015paradox}
Brandy~L Aven.
\newblock The paradox of corrupt networks: {An} analysis of organizational
  crime at {Enron}.
\newblock {\em Organization Science}, 26(4):980--996, 2015.

\bibitem{aven2019valley}
Brandy Aven, Lily Morse, and Alessandro Iorio.
\newblock The valley of trust: The effect of relational strength on monitoring
  quality.
\newblock {\em Organizational Behavior and Human Decision Processes}, 2019.

\bibitem{cheney2010contingent}
Dorothy~L Cheney, Liza~R Moscovice, Marlies Heesen, Roger Mundry, and Robert~M
  Seyfarth.
\newblock Contingent cooperation between wild female baboons.
\newblock {\em Proceedings of the National Academy of Sciences},
  107(21):9562--9566, 2010.

\bibitem{allen2008reciprocal}
Wesley Allen-Arave, Michael Gurven, and Kim Hill.
\newblock Reciprocal altruism, rather than kin selection, maintains nepotistic
  food transfers on an {Ache} reservation.
\newblock {\em Evolution and Human Behavior}, 29(5):305--318, 2008.

\bibitem{snijders2011statistical}
Tom~AB Snijders.
\newblock Statistical models for social networks.
\newblock {\em Annual Review of Sociology}, 37, 2011.

\bibitem{snijders1999social}
Tom~AB Snijders and David~A Kenny.
\newblock The social relations model for family data: A multilevel approach.
\newblock {\em Personal Relationships}, 6(4):471--486, 1999.

\bibitem{koster2019statistical}
Jeremy Koster, George Leckie, and Brandy Aven.
\newblock Statistical methods and software for the {Multilevel Social Relations
  Model}.
\newblock {\em Field Methods}, page 1525822X19889011, 2019.

\bibitem{koster2018effects}
Jeremy Koster and Brandy Aven.
\newblock The effects of individual status and group performance on network
  ties among teammates in the {National Basketball Association}.
\newblock {\em PloS one}, 13(4), 2018.

\bibitem{koster2014food}
Jeremy~M Koster and George Leckie.
\newblock Food sharing networks in lowland {Nicaragua:} an application of the
  social relations model to count data.
\newblock {\em Social Networks}, 38:100--110, 2014.

\bibitem{loncke2019social}
Justine Loncke, William~L Cook, Jenae~M Neiderhiser, and Tom Loeys.
\newblock {The Social Relations Model for Count Data}.
\newblock {\em Methodology}, 2019.

\bibitem{koster2018family}
Jeremy Koster.
\newblock Family ties: the multilevel effects of households and kinship on the
  networks of individuals.
\newblock {\em Royal Society Open Science}, 5(4):172159, 2018.

\bibitem{jorgensen2018using}
Terrence~D Jorgensen, K~Jean Forney, Jeffrey~A Hall, and Steven~M Giles.
\newblock Using modern methods for missing data analysis with the social
  relations model: A bridge to social network analysis.
\newblock {\em Social Networks}, 54:26--40, 2018.

\bibitem{koster2015multilevel}
Jeremy Koster, George Leckie, Andrew Miller, and Raymond Hames.
\newblock Multilevel modeling analysis of dyadic network data with an
  application to {Ye'kwana} food sharing.
\newblock {\em American Journal of Physical Anthropology}, 157(3):507--512,
  2015.

\bibitem{browne2005variance}
William~J Browne, Swamy~V Subramanian, Kelvyn Jones, and Harvey Goldstein.
\newblock Variance partitioning in multilevel logistic models that exhibit
  overdispersion.
\newblock {\em Journal of the Royal Statistical Society: Series A (Statistics
  in Society)}, 168(3):599--613, 2005.

\bibitem{snijders2011multilevel}
Tom~AB Snijders and Roel~J Bosker.
\newblock {\em Multilevel analysis: An introduction to basic and advanced
  multilevel modeling}.
\newblock Sage, 2011.

\bibitem{goldstein2011multilevel}
Harvey Goldstein.
\newblock {\em Multilevel statistical models}, volume 922.
\newblock John Wiley \& Sons, 2011.

\bibitem{mcelreath2020statistical}
Richard McElreath.
\newblock {\em Statistical rethinking: A {Bayesian} course with examples in {R
  and Stan}}.
\newblock CRC press, 2020.

\bibitem{finkel2008speed}
Eli~J Finkel and Paul~W Eastwick.
\newblock Speed-dating.
\newblock {\em Current Directions in Psychological Science}, 17(3):193--197,
  2008.

\bibitem{argote2018effects}
Linda Argote, Brandy~L Aven, and Jonathan Kush.
\newblock The effects of communication networks and turnover on transactive
  memory and group performance.
\newblock {\em Organization Science}, 29(2):191--206, 2018.

\bibitem{back2010social}
Mitja~D Back and David~A Kenny.
\newblock The social relations model: How to understand dyadic processes.
\newblock {\em Social and Personality Psychology Compass}, 4(10):855--870,
  2010.

\bibitem{krivitsky2009representing}
Pavel~N Krivitsky, Mark~S Handcock, Adrian~E Raftery, and Peter~D Hoff.
\newblock Representing degree distributions, clustering, and homophily in
  social networks with latent cluster random effects models.
\newblock {\em Social Networks}, 31(3):204--213, 2009.

\bibitem{sweet2018estimating}
Tracy~M Sweet and Qiwen Zheng.
\newblock Estimating the effects of network covariates on subgroup insularity
  with a hierarchical mixed membership stochastic blockmodel.
\newblock {\em Social Networks}, 52:100--114, 2018.

\bibitem{minhas2019inferential}
Shahryar Minhas, Peter~D Hoff, and Michael~D Ward.
\newblock Inferential approaches for network analysis: {AMEN} for latent factor
  models.
\newblock {\em Political Analysis}, 27(2):208--222, 2019.

\bibitem{zijlstra2017regression}
Bonne~JH Zijlstra.
\newblock Regression of directed graphs on independent effects for density and
  reciprocity.
\newblock {\em The Journal of Mathematical Sociology}, 41(4):185--192, 2017.

\bibitem{butts20084}
Carter~T Butts.
\newblock A relational event framework for social action.
\newblock {\em Sociological Methodology}, 38(1):155--200, 2008.

\bibitem{stadtfeld2017interactions}
Christoph Stadtfeld and Per Block.
\newblock Interactions, actors, and time: Dynamic network actor models for
  relational events.
\newblock {\em Sociological Science}, 4:318--352, 2017.

\bibitem{leckie2014modeling}
George Leckie, Robert French, Chris Charlton, and William Browne.
\newblock Modeling heterogeneous variance--covariance components in two-level
  models.
\newblock {\em Journal of Educational and Behavioral Statistics},
  39(5):307--332, 2014.

\end{thebibliography}

\end{document}